\documentclass[12pt]{iopart}

\usepackage{iopams}  
\usepackage{graphicx}
\begin{document}

\title[Deformation in amorphous-crystalline nanolaminates]{Deformation in amorphous-crystalline nanolaminates -- an effective-temperature theory and interaction between defects}

\author{Charles K C Lieou$^1$, Jason R Mayeur$^2$ and Irene J Beyerlein$^{2,3}$}
\address{$^1$ Earth and Environmental Sciences and Center for Nonlinear Studies, Los Alamos National Laboratory, Los Alamos, NM 87545, USA}
\ead{clieou@lanl.gov}
\address{$^2$ Theoretical Division, Los Alamos National Laboratory, Los Alamos, NM 87545, USA}
\ead{jmayeur@lanl.gov}
\address{$^3$ Mechanical Engineering and Materials Departments, University of California, Santa Barbara, CA 93106, USA}
\ead{beyerlein@engineering.ucsb.edu}

\vspace{10pt}
\begin{indented}
\item[]November 2016
\end{indented}

\begin{abstract}
Experiments and atomic-scale simulations suggest that the transmission of plasticity carriers in deforming amorphous-crystalline nanolaminates is mediated by the biphase interface between the amorphous and crystalline layers. In this paper, we present a micromechanics model for these biphase nanolaminates that describes defect interactions through the amorphous-crystalline interface (ACI). The model is based on an effective-temperature framework to achieve a unified description of the slow, configurational atomic rearrangements in both phases when driven out of equilibrium. We show how the second law of thermodynamics constrains the density of defects and the rate of configurational rearrangements, and apply this framework to dislocations in crystalline solids and shear transformation zones (STZs) in amorphous materials. The effective-temperature formulation enables us to interpret the observed movement of dislocations to the ACI and the production of STZs at the interface as a ``diffusion'' of configurational disorder across the material. We demonstrate favorable agreement with experimental findings reported in (Kim et al., Adv. Funct. Mater., 2011), and demonstrate how the ACI acts as a sink of dislocations and a source of STZs.
\end{abstract}

%\maketitle

\section{Introduction}

\begin{figure}
\begin{center}
\includegraphics[scale=0.6]{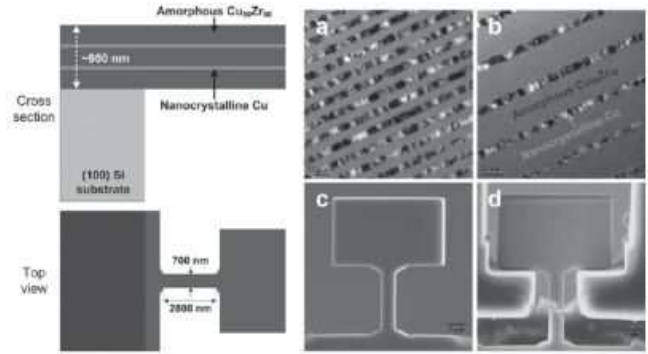}
\caption{\label{fig:setup}CuZr/Cu amorphous-crystalline nanolaminate sample undergoing tensile testing. (a, b) Bright-field TEM images for cross-sections of the nanolaminates with amorphous layer thickness $h_a = $ (a) 17 nm and (b) 128 nm; the crystalline layer thickness is $h_c = 16$ nm. (c, d) SEM images of the freestanding tensile samples (c) before and (d) after tension. Adapted with permission from \cite{kim_2011}.}
\end{center}
\end{figure}

Amorphous-crystalline nanolaminates are heterogeneous structures fabricated by alternately stacking nano-thick layers of nanocrystalline materials (commonly nanocrystalline copper) and amorphous materials (often metallic glasses) upon one another~\cite{wang_2007,arman_2011,kim_2011,brandl_2013,zhang_2013}, often by means of magnetron sputtering. Figure \ref{fig:setup} shows a CuZr/Cu amorphous-crystalline nanolaminate subject to uniaxial loading. Experiments have shown that the addition of metallic glass layers greatly enhances the strength and ductility of the nanocrystalline material~\cite{wang_2007}, and that this effect is especially pronounced when the metallic glass thickness is below some threshold~\cite{kim_2011,arman_2011}. While some microscopy imaging and molecular dynamics (MD) simulations suggest that the exceptional strength and ductility may be accounted for by the inhibitory effect of the metallic glass layer on shear band propagation in the crystalline layer~\cite{wang_2007}, others suggest that the crystalline layer obstructs shear band propagation in the metallic glass~\cite{zhang_2013}. In either case, the amorphous-crystalline interface (ACI) mediates the plastic interaction between the two constituent materials. The ACI absorbs plasticity carriers -- dislocations in the crystalline material and shear transformation zones (STZs) in the amorphous material -- coming in from one side and triggers emission into the other, thereby playing an important role in controlling the deformation of the heterogeneous nanolayered structure.

A predictive description that links the physical mechanisms underlying these observations in atomic-scale simulation to the macroscopic deformation response would be useful in understanding their behavior and ultimately designing their microstructures. However, the classes of models that treat amorphous materials alone and crystalline solids alone, in practice, use different formulations or frameworks. The commonly used theories of dislocations in crystalline solids are based on the thermodynamics of slip overcoming obstacles (e.g.,~\cite{follansbee_1988,kocks_2003}). 

Those that describe the development of STZs often directly address the flow of energy and entropy in the deforming material or the principle of symmetry. It is unclear whether the basis of defect theories of two dissimilar materials can be chosen independently and still provide a reliable mechanics model. Adopting a generalized framework, in which the kinetics of both types of defects can be described, ought to be the more straightforward path to take. Accordingly, for the present ACI system of interest, a unified description of the dynamics of dislocations and STZs would be needed to elucidate the role of the ACI, and to predict how the heterogeneous multilayers would deform and ultimately fail. The aim of this work is to model the complex interactions between the plasticity carriers that arise when these two very different types of materials are joined together in an ACI system:  STZs in the amorphous layer and dislocations in the crystalline layer. 

There is consensus in the literature that shear transformation zones (STZs) are the plasticity carriers in amorphous materials such as metallic glasses, colloids, and foams~\cite{falk_1998,langer_2008,falk_2011,wang_2007,arman_2011,kamrin_2014}. STZs are localized clusters of atoms or molecules susceptible to nonaffine, irreversible rearrangement under an applied stress. The STZ population is controlled by an effective temperature that pertains to the slow, atomic configurational degrees of freedom of the deforming material which fall out of equilibrium with the thermal background. The theory has been invoked to explain the yielding transition~\cite{langer_2008,langer_2015a}, shear banding~\cite{manning_2007,manning_2009}, and crack propagation and fracture~\cite{rycroft_2012} in metallic glasses. The deformation in a crystalline solid, with dislocations as the plasticity carriers, also involves infrequent (relative to atomic vibrations), configurational atomic rearrangements analogous to those occurring in a metallic glass. The effective temperature that describes those configurational degrees of freedom, as well as how energy in the form of external work flows through those degrees of freedom and results in plastic deformation, must also play an important role. Recently Langer presented an effective-temperature dislocation theory~\cite{langer_2010,langer_2015b}. Such an effective-temperature theory of dislocations would be a viable path toward a unified description of interactions between STZs and dislocations in different materials, such that those occurring in a deforming amorphous-crystalline nanolaminate, where we already know that the effective temperature controls the plasticity in the amorphous metallic glass layers.

%In contrast, most of the commonly used dislocation theories~\cite{follansbee_1988,kocks_2003} make rather phenomenological assumptions about the connection between strain rate and the rate at which dislocations are created, and attempt power-law fits to experimental data, with no reference to the effective temperature.

The rest of this paper is structured as follows. We present a simplified discussion of the concept of effective temperature in section \ref{sec:2}, highlighting how the thermodynamic principles of energy conservation and nondecreasing entropy constrain the defect densities and their internal dynamics. Next, in sections \ref{sec:3} and \ref{sec:4}, we review the effective-temperature theories of STZs in metallic glasses, and dislocations in crystalline solids, and deduce the evolution equations for the internal state variables from simple physical principles and basic assumptions. Then, we discuss in section \ref{sec:5} how the movement of the plasticity carriers across the layered material can be interpreted as a flow of configurational disorder governed by a diffusive term for the effective temperature. We demonstrate in section \ref{sec:6} that the evolution equations for the internal variables, along with a few physically well-defined parameters, produce close agreement with the experimental results of Kim, Jang and Greer \cite{kim_2011}.

\section{The effective temperature}\label{sec:2}

This section provides a brief introduction to the concept of effective temperature. In this discussion we will make no reference to the nature of the structural flow defects (dislocations or STZs), or that of the deforming material, but shall show that important conclusions about the flow defect densities can be drawn in a systematic and straightforward manner. For a full review of effective-temperature thermodynamics the reader is referred to~\cite{bouchbinder_2009a,bouchbinder_2009b,bouchbinder_2009c}.

In a deforming solid, the atomic configurational degrees of freedom -- those that describe their interactions and relative positions -- are driven out of equilibrium with the fast, kinetic-vibrational degrees of freedom by external forces. The configurational and kinetic-vibrational subsystems are only weakly coupled to one another. Despite this, the two subsystems do exchange energy with one another when clusters of atoms rearrange in a nonaffine, irreversible manner, albeit extremely slowly when compared to the atomic time scale of order $\tau \sim 10^{-12}$ s. Thus we focus exclusively on the configurational subsystem; denote by $U_C$ and $S_C$ its energy and entropy. $U_C$ is a function of $S_C$ and the density $\rho$ of defects, and perhaps of other order parameters that we omit for the time being. Conversely, $S_C$ is the configurational entropy computed by counting the number of atomic configurations, or the number of possible arrangements of defects, at fixed energy $U_C$ and defect density $\rho$. Next, define the effective temperature (here with units of energy):
\begin{equation}\label{eq:Teff_def}
 \chi = \left( \frac{\partial U_C}{\partial S_C} \right)_{\rho} .
\end{equation}

In a deforming solid undergoing atomic rearrangements, defects are driven by external forces to explore a large swath of the available configurational phase space; thus, the configurational degrees of freedom must be maximizing the entropy $S_C$ during this process. The configurational energy $U_C$, meanwhile, is determined by the balance between the external work rate and the rate at which energy is dissipated into kinetic-vibrational subsystem. This process minimizes the configurational free energy given by
\begin{equation}
 F_C = U_C - \chi S_C .
\end{equation}
If $e_D$ is the characteristic energy of a single defect, and $\rho$ is the population density of defects, then in the noninteracting defects approximation, $U_C = V e_D \rho$, where $V$ is the total volume of the solid. Meanwhile, elementary statistical mechanics shows that $S_C \propto - \rho \ln \rho + \rho$. Minimizing $F_C$, we find that the instantaneous, steady-state defect density $\rho^{\rm{ss}}$ must be given by
\begin{equation}
 \rho^{\rm{ss}} \propto e^{- e_D / \chi} .
\end{equation}
This argument applies directly to dislocations in a crystalline solid, with the caveat that when we refer to the dislocation density per unit area, we replace the volume $V$ by some cross-sectional area $A$. For STZs there is an extra order parameter $m$ that denotes the orientation relative to the applied stress (see section \ref{sec:4} below); to a good approximation STZs exist in two states for each given stress configuration. Thus the preceding result acquires an extra factor of two when carried over to the STZ density $\Lambda$, here normalized over the total number of atomic sites:
\begin{equation}
 \Lambda^{\rm{ss}} = 2 e^{-e_Z / \chi} ,
\end{equation}
where $e_Z$ is a characteristic STZ formation energy. Thus, the evolution of the defect densities is largely controlled by the temporal evolution of the effective temperature, governed by the input power and the energy dissipation rate.

Next, we proceed to derive the evolution equation for the effective temperature $\chi$ from the first law of thermodynamics as follows. Let $U_K$ and $S_K$ denote the energy and entropy, respectively, of the kinetic-vibrational degrees of freedom, and denote by $\theta$ the thermal temperature in energy units. Then the total energy equals $U_K + U_C$, and the energy balance equation reads
\begin{equation}\label{eq:first_law}
 \dot{U}_C + \dot{U}_K = V \sigma \dot{\epsilon}^{\rm{pl}} = \chi \dot{S}_C + \left( \frac{\partial U_C}{\partial \rho} \right)_{S_C} \dot{\rho} + \theta \dot{S}_K .
\end{equation}
In (\ref{eq:first_law}), $\sigma$ and $\dot{\epsilon}^{\rm{pl}}$ denote the stress and plastic strain rate. From here onwards, we specialize to the case of tensile deformation, so that it is permissible in most cases to use the magnitudes $\sigma$ and $\dot{\epsilon}^{\rm{pl}}$ instead of writing the tensorial product for the input power $V \boldsymbol{\sigma} : \dot{\boldsymbol{\epsilon}}^{\rm{pl}} $ in full. (Note that only the plastic work of deformation plays a role; the elastic work of deformation cancels out of this equation following the argument in \cite{bouchbinder_2009b}.) Meanwhile, the second law of thermodynamics says that $\dot{S}_C + \dot{S}_K \geq 0$; eliminating $S_C$ using (\ref{eq:first_law}), we find
\begin{equation}
 V \sigma \dot{\epsilon}^{\rm{pl}} - \left( \frac{\partial U_C}{\partial \rho} \right)_{S_C} \dot{\rho} + (\chi - \theta) \dot{S}_K \geq 0 .
\end{equation}
Since this holds for all possible motions of the state variables, each independent term must separately be non-negative. Because $V \sigma \dot{\epsilon}^{\rm{pl}} \geq 0$ for all practical purposes, it follows that
\begin{eqnarray}
 - \left( \frac{\partial U_C}{\partial \rho} \right)_{S_C} \dot{\rho} & \geq & 0 ; \\
 (\chi - \theta) \dot{S}_K & \geq & 0 .
\end{eqnarray}
In order for each of these inequalities to hold, both multiplicative factors in the inequality must switch signs at the same point. Thus the first inequality constrains the steady-state defect density $\rho^{\rm{ss}}$ in the same manner as before. (In the case of STZs, for which there is an extra orientational order parameter, an extra constraint for the transition rate arises; see for example \cite{bouchbinder_2009c}.) The second inequality says that $\dot{S}_K$ must be proportional to the temperature difference $\chi - \theta$, which must be non-negative.  We write this in the form
\begin{equation}
 \theta \dot{S}_K = {\cal K}(\chi, \theta) (\chi - \theta) ,
\end{equation}
where ${\cal K}$ is a non-negative coupling coefficient between the configurational and kinetic-vibrational subsystems. Finally, substituting this into (\ref{eq:first_law}), and using $\chi \dot{S}_C \simeq V c^{\rm{eff}} \dot{\chi}$, where $c^{\rm{eff}}$ is an effective specific heat capacity, we arrive at the evolution equation for the effective temperature:
\begin{equation}\label{eq:efftemp_evolution}
 V c^{\rm{eff}} \dot{\chi} = V \sigma \dot{\epsilon}^{\rm{pl}} - {\cal K}(\chi, \theta) (\chi - \theta) - \left( \frac{\partial U_C}{\partial \rho} \right) \dot{\rho} .
\end{equation}
We shall now apply this effective-temperature formulation to STZs in amorphous solids and dislocations in crystalline solids independently.

\section{Effective-temperature theory of STZs in amorphous solids}\label{sec:3}

This section provides a brief review of the STZ theory of plastic deformation in amorphous solids. The interested reader is referred to, for example, \cite{langer_2008,falk_2011,bouchbinder_2009c}, for details and derivations.

In the STZ description, the order parameters of interest are the STZ density $\Lambda$ and the orientational bias $m$. STZs fluctuate into and out of existence due to the thermal motion of the atoms and the mechanical work input. The former is unimportant if we confine ourselves to metallic glasses below the glass transition temperature as in \cite{wang_2007,kim_2011}, while the latter is described by a mechanical noise strength $\Gamma$. When subjected to external stresses, the atoms in an STZ undergo irreversible arrangements and produce plastic strain. The tensorial relation between the plastic strain rate and the stress is
\begin{equation}
 \tau \dot{\gamma}^{\rm{pl}}_{ij} = \epsilon_0 {\cal C}( \bar{s} ) \Lambda \frac{s_{ij}}{\bar{s}} \left( {\cal T}(\bar{s}) - m \right).
\end{equation}
Here, $\tau$ is the fundamental molecular time scale or the inverse attempt frequency, and $\epsilon_0$ is the ratio of the STZ plastic core volume to the atomic volume. ${\cal C}(\bar{s})$ and ${\cal T}(\bar{s})$ are symmetric and antisymmetric combinations of the forward and backward STZ transition rates ${\cal R}(\pm \bar{s})$:
\begin{equation}
 {\cal C}(\bar{s}) \equiv \frac{1}{2} \left( {\cal R}(\bar{s}) + {\cal R}(- \bar{s}) \right) ; \qquad {\cal T}(\bar{s}) \equiv \frac{{\cal R}(\bar{s}) - {\cal R}(- \bar{s})}{{\cal R}(\bar{s}) + {\cal R}(- \bar{s})} ,
\end{equation}
where $\bar{s} \equiv \sqrt{\frac{1}{2} s_{ij} s_{ij}}$, and $s_{ij}$ is the deviatoric stress tensor. In a similar vein the plastic strain rate tensor and the deviatoric plastic strain rate are related through $\bar{\dot{\gamma}}^{\rm{pl}} \equiv \sqrt{\frac{1}{2} \dot{\gamma}^{\rm{pl}}_{ij} \dot{\gamma}^{\rm{pl}}_{ij}}$. Thus, in the case of tensile loading, if we choose coordinate systems such that the only nonzero element of the \textit{total} stress tensor is $\sigma_{xx} = \sigma$, and that the corresponding plastic strain rate is $\dot{\epsilon}^{\rm{pl}}$, then the nonzero elements of the deviatoric stress tensor $s_{ij} = \sigma_{ij} - \frac{1}{3} \rm{tr} (\sigma_{ij})$ are
\begin{equation}\label{eq:dev_s}
 s_{xx} = \frac{2}{3} \sigma; \qquad s_{yy} = s_{zz} = - \frac{1}{3} \sigma,
\end{equation}
and the deviatoric plastic strain tensor has nonzero elements
\begin{equation}\label{eq:dev_g}
 \dot{\gamma}^{\rm{pl}}_{xx} = \dot{\epsilon}^{\rm{pl}}; \qquad \dot{\gamma}^{\rm{pl}}_{yy} = \dot{\gamma}^{\rm{pl}}_{zz} = - \dot{\epsilon}^{\rm{pl}}/2 .
\end{equation}
Thus, $\sigma = \sqrt{3} \bar{s}$ and $\dot{\epsilon}^{\rm{pl}} = (2 / \sqrt{3}) \bar{\dot{\gamma}}^{\rm{pl}}$. The plastic work of deformation, or the dissipation rate excluding those attributed to the change of internal state variables, is $\sigma \dot{\epsilon}^{\rm{pl}} = 2 \bar{s} \bar{\dot{\gamma}}^{\rm{pl}}$.

The rest of the paper is devoted to tensile deformation. For convenience, from now on we use the experimentally measured tensile stress $\sigma$ instead of the deviatoric stress $\bar{s}$ in the arguments for the STZ transition rate factors ${\cal C}(\bar{s})$ and ${\cal T}(\bar{s})$. The tensile stress evolves with time according to linear elasticity; that is,
\begin{equation}\label{eq:a_sigma}
 \dot{\sigma} = E (\dot{\epsilon} - \dot{\epsilon}^{\rm{pl}} ),
\end{equation}
where $E$ is the Young modulus of the amorphous material, and $\dot{\epsilon}$ is the applied strain rate. The plastic strain rate $\dot{\epsilon}^{\rm{pl}}$ evolves according to
\begin{equation}\label{eq:a_vpl}
 \tau \dot{\epsilon}^{\rm{pl}} = \frac{2}{\sqrt{3}} \epsilon_0 \Lambda {\cal C} (\sigma) \left( {\cal T} (\sigma) - m \right) ,
\end{equation}
where, below the glass transition temperature \cite{bouchbinder_2009c}, as is the case in the experiments of interest \cite{wang_2007,kim_2011},
\begin{equation}\label{eq:a_Cs}
 {\cal C}(\sigma) = \exp \left( - \frac{T_E}{T} \right) \cosh \left( \frac{\epsilon_0 \sigma a^3}{\sqrt{3} \chi} \right) ,
\end{equation}
and
\begin{equation}\label{eq:a_Ts}
 {\cal T}(\sigma) = \tanh \left( \frac{\epsilon_0 \sigma a^3}{\sqrt{3} \chi} \right); \qquad 
 m = \cases{
 {\cal T}(\sigma) & if $\sigma {\cal T}(\sigma) \leq \sigma_0$ \\
 \sigma_0 / \sigma & if $\sigma {\cal T}(\sigma) > \sigma_0$ . \\
 }
\end{equation}
Here, $T_E$ is an activation temperature and $a$ is the atomic radius. The stress $\sigma_0$ may be interpreted as a yield stress parameter. It emerges from the proportionality between the mechanical noise strength $\Gamma$ and the plastic dissipation per STZ as a proportionality constant \cite{bouchbinder_2009c}:
\begin{equation}
 \Gamma = \frac{\sqrt{3} \tau \sigma \dot{\epsilon}^{\rm{pl}}}{\epsilon_0 \sigma_0 \Lambda} .
\end{equation}

The STZ density $\Lambda$ evolves according to the equation
\begin{equation}\label{eq:a_Lambda}
 \dot{\Lambda} = \frac{\Gamma}{\tau} (2 e^{- e_Z / \chi} - \Lambda) .
\end{equation}
As before, the quantity $e_Z$ is the STZ formation energy, and in section \ref{sec:2}, we argued that at steady state $\Lambda^{\rm{ss}} = 2 e^{- e_Z / \chi}$. Finally, the effective temperature $\chi$ evolves according to (\ref{eq:efftemp_evolution}). After some algebraic simplifications, we find 
\begin{equation}\label{eq:a_chi}
 c^{\rm{eff}} \dot{\chi} = \sigma \dot{\epsilon}^{\rm{pl}} \left( 1 - \frac{\chi}{\chi_0} \right) - \frac{e_Z}{a^3} \dot{\Lambda} .
\end{equation}
In (\ref{eq:a_chi}), $c^{\rm{eff}}$ is the so-called effective heat capacity. It has the dimensions of inverse volume. The effective temperature evolves to some constant value $\chi_0$ in the steady state. Strictly speaking, the steady-state value should be a function of the strain rate. At strain rates slower than the internal relaxation rate controlled by the atomic vibration frequency (i.e., when $\tau \dot{\gamma}^{\rm{pl}} \ll 1$); however, the approximation of a constant $\chi_0$ is sufficient.

Note that the equation of motion for $\Lambda$, (\ref{eq:a_Lambda}), does not contain an overall factor of $\Lambda$, which we have assumed to be small. On the other hand, the equation of motion for $\chi$, (\ref{eq:a_chi}), contains the small factor of $\Lambda$ through $\dot{\epsilon}^{\rm{pl}}$. Thus $\Lambda$ is a fast variable while $\chi$ is a slow variable. For most purposes, we can use the steady-state approximation $\Lambda \approx \Lambda^{\rm{ss}} = 2 e^{-e_Z / \chi}$, which we shall do from here onwards.

Summarizing, the equation of motion in the amorphous layer are given by (\ref{eq:a_sigma}), (\ref{eq:a_chi}), (\ref{eq:a_vpl}), and (\ref{eq:a_Lambda}):
\begin{eqnarray}
 \dot{\sigma} &=& E (\dot{\epsilon} - \dot{\epsilon}^{\rm{pl}} ), \\
 c^{\rm{eff}} \dot{\chi} &=& \sigma \dot{\epsilon}^{\rm{pl}} \left( 1 - \frac{\chi}{\chi_0} \right) , \\
 \tau \dot{\epsilon}^{\rm{pl}} &=& \frac{4}{\sqrt{3}} \epsilon_0 e^{-e_Z/\chi} {\cal C} (\sigma) \left( {\cal T} (\sigma) - m \right) .
\end{eqnarray}

\section{Effective-temperature theory of dislocations in crystalline solids}\label{sec:4}

Deformation in the crystalline layer is mediated by dislocations. Like STZs, the motion of dislocations can be analyzed in a statistical-thermodynamic framework. The development here closely follows that of \cite{langer_2010,langer_2015b}. As in the amorphous layer, the tensile stress increases linearly with the elastic strain rate; thus,
\begin{equation}
 \dot{\sigma} = E ( \dot{\epsilon} - \dot{\epsilon}^{\rm{pl}} ) .
\end{equation}
Note, however, that both the Young's modulus $E$ and the plastic strain rate $\dot{\epsilon}^{\rm{pl}}$ generally differ from those in the amorphous layer.

The derivation of the expression for the plastic strain rate starts with the Orowan relation for the plastic shear rate $\dot{\gamma}^{\rm{pl}}$:
\begin{equation}\label{eq:c_gpl1}
 \dot{\gamma}^{\rm{pl}} = \rho b v .
\end{equation}
In this equation, $\rho$ is the areal density of mobile dislocations, to be distinguished from the volume density or number density in section \ref{sec:2} above. (Here, we do not study the motions of individual dislocations; rather, we apply coarse-graining and use a dislocation density description.) The quantity $b$ is the length of the Burgers vector, and $v = l / \tau_P (\sigma)$ is the average speed at which dislocations move in the crystal, expressed in terms of average spacing $l = 1 / \sqrt{\rho}$ between dislocations, and the depinning rate $1 / \tau_P (\sigma)$. Depinning is a thermally activated process with an assumed stress-dependent barrier of the form
\begin{equation}
 U_P (\sigma) = k_B T_P e^{-s / \sigma_T},
\end{equation}
where $s$ is the shear stress $\sigma_T$ is  the Taylor (depinning) stress
\begin{equation}\label{eq:c_st}
 \sigma_T = \mu_T b \sqrt{\rho},
\end{equation}
with $\mu_T$ being an effective shear modulus on the order of $1 / 30$ times the shear modulus $\mu$. As such, the depinning rate is
\begin{equation}
 \frac{1}{\tau_P (s)} = \frac{1}{\tau}f_P (s) ,
\end{equation}
where
\begin{equation}
 f_P (s) = \exp \left( - \frac{T_P}{T} e^{-s / \sigma_T} \right).
\end{equation}
Then, the plastic strain rate, which must change sign as the stress direction is reversed, is
\begin{equation}\label{eq:c_gpl2}
 \dot{\gamma}^{\rm{pl}} = \frac{\sqrt{\bar{\rho}}}{\tau} [ f_P (s) - f_P (- s) ],
\end{equation}
where $\bar{\rho} \equiv b^2 \rho$ is a non-dimensional dislocation density.  The second term on the RHS of (\ref{eq:c_gpl2}) accounts for reverse transitions; it is typically neglected in practice, and will be dropped in the following.

To convert these expressions to a form appropriate for describing tensile deformation, we first rewrite equations (\ref{eq:c_gpl1}) and (\ref{eq:c_gpl2}) using the deviatoric stress and plastic strain rate tensors $s_{ij}$ and $\dot{\gamma}^{\rm{pl}}_{ij}$, and the stress and strain rate invariants $\bar{s}$ and $\bar{\dot{\gamma}}^{\rm{pl}}$, as in section \ref{sec:3}. Thus, the Orowan relation, (\ref{eq:c_gpl1}), becomes
\begin{equation}
 \dot{\gamma}^{\rm{pl}}_{ij} = \frac{\rho}{2} \frac{s_{ij}}{\bar{s}} b v .
\end{equation}
(This reduces directly to (\ref{eq:c_gpl1}) in the case of simple shear, for which the only nonvanishing elements of the stress and strain rate tensors are $s_{xy} = s_{yx} = s$ and $\dot{\gamma}^{\rm{pl}}_{xy} = \dot{\gamma}^{\rm{pl}}_{yx} = \dot{\gamma}^{\rm{pl}}/2$.) For tensile deformation, use of (\ref{eq:dev_s}) and (\ref{eq:dev_g}) for the nonzero elements of the deviatoric plastic stress and strain rate tensors gives
\begin{equation}
 \dot{\epsilon}^{\rm{pl}} = \frac{\rho}{2} \frac{2 \sigma / 3}{\sigma / \sqrt{3}} v = \frac{1}{\sqrt{3}} \rho b v ,
\end{equation}
so that (\ref{eq:c_gpl2}) becomes
\begin{equation}\label{eq:c_vpl}
 q \equiv \tau \dot{\epsilon}^{\rm{pl}} = \sqrt{\tilde{\rho}}  f_P (\bar{\sigma}),
\end{equation}
where now $\tilde{\rho} = \bar{\rho} / 3$ and $f_P$ is now expressed as function of the von Mises effective stress $\bar{\sigma} = \sqrt{3}\bar{s}$, i.e.
\begin{equation}
 f_P (\bar{\sigma}) = \exp \left( - \frac{T_P}{T} e^{-\bar{\sigma} / \sigma_T} \right).
\end{equation}

The dislocation density evolves according the second law of thermodynamics.  Following the analysis in \cite{langer_2015b}, it approaches some steady state $\rho^{\rm{ss}} (\chi) = (1 / a^2) e^{- e_D / \chi}$, controlled by the effective temperature $\chi$, with $e_D$ being the energy per dislocation. The rate at which $\rho$ approaches $\rho^{\rm{ss}} (\chi)$ is assumed to be proportional to the rate of plastic work, and inversely proportional to the dislocation energy per unit length $\gamma_D$. Thus,
\begin{equation}\label{eq:c_rho}
 \dot{\rho} = \kappa_{\rho} \frac{\sigma \dot{\epsilon}^{\rm{pl}}}{\gamma_D} \left[ 1 - \frac{\rho}{\rho^{\rm{ss}} (\chi)} \right] ,
\end{equation}
where $\kappa_{\rho}$ is a dimensionless conversion factor that determines the fraction of energy input that is converted into dislocations.

Meanwhile, the equation for the effective temperature describes the flow of entropy and, as in the amorphous case, is a statement of the first law of thermodynamics:
\begin{equation}\label{eq:c_chi}
 c^{\rm{eff}} \dot{\chi} = \sigma \dot{\epsilon}^{\rm{pl}} \left( 1 - \frac{\chi}{\chi_0} \right) - \gamma_D \dot{\rho}.
\end{equation}

To proceed, first assume the normalization $b = \sqrt{3} a$, or $\tilde{\rho} = a^2 \rho$, for the dislocation density $\rho$. Next, note that (\ref{eq:c_vpl}) can be solved explicitly for the stress as a function of the strain rate and the dislocation density:
\begin{equation}
 \frac{\sigma}{\sigma_T} = \ln \left( \frac{T_P}{T} \right) - \ln \left[ \ln \left( \frac{\sqrt{\tilde{\rho}}}{q} \right) \right] \equiv \nu (T, \tilde{\rho}, q).
\end{equation}
Here we have taken advantage of the fact that $\bar{\sigma} = \sigma$ under unaxial loading conditions.  Because the elastic modulus $E$ is much larger than the other stress scales in the problem, we make use of the approximation $\dot{\epsilon}^{\rm{pl}} \approx \dot{\epsilon}$, or $q \approx q_0 \equiv \tau \dot{\epsilon}$. As such, the only dynamical equations would concern the effective temperature $\chi$ and the normalized dislocation density $\tilde{\rho}$. Their equations of motion are
\begin{eqnarray}
 \label{eq:c_chi2} c^{\rm{eff}} \dot{\chi} &=& \sigma \dot{\epsilon}^{\rm{pl}} \left( 1 - \frac{\chi}{\chi_0} \right) - \gamma_D \frac{ \dot{\tilde{\rho}}}{a^2}, \\
 \label{eq:c_rho2} \dot{\tilde{\rho}} &=& \kappa_{\rho} a^2 \frac{\sigma \dot{\epsilon}^{\rm{pl}}}{\gamma_D} \left[ 1 - \frac{\tilde{\rho}}{e^{- e_D / \chi}} \right] .
\end{eqnarray}
The tensile stress is directly given by
\begin{equation}\label{eq:c_sigma2}
 \sigma = \bar{\mu}_T \sqrt{\tilde{\rho}} \, \nu(T, \tilde{\rho}, q_0),
\end{equation}
where $\bar{\mu}_T$ is proportional to the reduced shear modulus $\mu_T$, defined above in (\ref{eq:c_st}): $\bar{\mu}_T = \sqrt{3} \mu_T$.

We close this section with some comments on the dimensionless conversion factor $\kappa_{\rho}$ in (\ref{eq:c_rho2}), which determines the fraction of input power that is stored in the form of dislocations. To understand the physics behind this parameter, we consider the onset of strain hardening, when $q = q_0$ but the dislocation density $\tilde{\rho}$ is still small and has yet to reach its steady-state value. The stress at the onset of hardening is simply the Taylor stress, so that from (\ref{eq:c_rho2}), we get
\begin{equation}
 \left( \frac{d \tilde{\rho}}{d \epsilon} \right)_{\rm{onset}} \approx \frac{\kappa_{\rho} a^2 \sigma_T}{\gamma_D} = \frac{\kappa_{\rho} a^2 \bar{\mu}_T}{\gamma_D} \sqrt{\tilde{\rho}} .
\end{equation}
This can be substituted into (\ref{eq:c_sigma2}) to give
\begin{equation}\label{eq:onset}
 \left( \frac{d \sigma}{d \epsilon} \right)_{\rm{onset}} \approx \left( \frac{d \sigma_T}{d \epsilon} \right)_{\rm{onset}} = \frac{\kappa_{\rho} \bar{\mu}_T^2 a^2}{2 \gamma_D}.
\end{equation}
However, if we directly use the full versions of equations (\ref{eq:c_rho2}) and (\ref{eq:c_sigma2}) to compute the onset rate, we get an extra factor $\nu (T, \tilde{\rho}, q_0 )^2$ multiplying $\kappa_{\rho}$ on the right-hand side of (\ref{eq:onset}). Thus we conclude that
\begin{equation}\label{eq:kappa}
 \kappa_{\rho} = \frac{\tilde{\kappa}_{\rho}}{\nu (T, \tilde{\rho}, q_0 )^2},
\end{equation}
where $\tilde{\kappa}_{\rho}$ is a constant of order unity. Then, after some algebra, the evolution equation for $\tilde{\rho}$ becomes
\begin{equation}\label{eq:c_rho3}
 \dot{\tilde{\rho}} = \kappa_1 \frac{\sqrt{\tilde{\rho}} q_0}{\nu (T, \tilde{\rho}, q_0)} \left(1 - \frac{\tilde{\rho}}{e^{- e_D / \chi}} \right),
\end{equation}
where the constant
\begin{equation}
 \kappa_1 \equiv \tilde{\kappa}_{\rho} \frac{a^2 \bar{\mu}_T}{\gamma_D}
\end{equation}
is of order unity.

\section{Coupled amorphous-crystalline layers -- interaction between STZs and dislocations}\label{sec:5}

We are now in a position to combine the effective-temperature descriptions of STZs and dislocations from sections \ref{sec:3} and \ref{sec:4}, and model the interaction between the dislocations in the crystalline layers and the STZs in the amorphous layers in simple terms. From now on, we use the subscripts $a$ and $c$ to denote the quantities relevant to the amorphous and crystalline layers, respectively. Under isostrain conditions in the two constituents, the experimentally measured tensile stress is
\begin{equation}\label{eq:stress}
 \sigma \equiv \frac{\sigma_a h_a + \sigma_c h_c}{h_a + h_c} ,
\end{equation}
where $h_a$ and $h_c$ denote the layer thickness of the amorphous and crystalline layers, respectively. The assumption of co-deformation (isostrain) also implies that $\dot{\epsilon}_a = \dot{\epsilon}_c \equiv \dot{\epsilon}$, and in general $\dot{\epsilon}^{\rm{pl}}_a \neq \dot{\epsilon}^{\rm{pl}}_c$, and $\sigma_a$ does not necessarily equal $\sigma_c$.

Experiments and simulations (e.g.~\cite{wang_2007}) indicate that the amorphous-crystalline interface (ACI) acts as a sink of dislocations; an arriving dislocation from the crystalline layer gets absorbed and triggers an STZ that moves into the amorphous layer. Other studies (e.g.~\cite{zhang_2013}) seem to suggest that the stress concentration of an STZ near the ACI may be accommodated locally by the emission of a dislocation or an array of dislocations that moves into the f, which is also a plausible scenario. One way to interpret these dislocation/STZ interactions is through the lens of effective-temperature dynamics and the flow of entropy. Specifically, if the effective temperature of the amorphous layer somehow increases more slowly than in the crystalline layer during the deformation process, it is possible for entropy to flow from the crystalline layer to the amorphous layer, or for the effective temperature to ``diffuse'' into the amorphous layer. This entropy flow is manifested by the movement of dislocations in the crystalline layer into the amorphous-crystalline interface to trigger STZs that move into the amorphous layer. The opposite movement may occur if the effective temperature of the amorphous layer increases more quickly, and stays above that of the crystalline layer. In either case, the diffusion term that describes this process is of the form
\begin{equation}\label{eq:diff}
 \left( \frac{d \chi}{d t} \right)_{\rm{diff}} = D_0 a^2 \dot{\epsilon}^{\rm{pl}} \frac{\partial^2 \chi}{\partial y^2} ,
\end{equation}
where $\chi = \chi_c$ or $\chi_a$, $D_0 = D_c$ or $D_a$, where the conduction coefficients $D_c$ and $D_a$ in the two layers need not be equal, and $y$ is the spatial coordinate in the direction normal to the interface. A diffusion term of this type, proportional to the divergence of the ``configurational heat flux'', was invoked elsewhere~\cite{manning_2007,manning_2009,kamrin_2014} to model the shear-banding instability.

With equation (\ref{eq:diff}) in mind, and using the total strain $\epsilon$ as the independent variable in the dynamical equations, the equations of motion for the coupled amorphous-crystalline nanolaminate are
\begin{eqnarray}
 \label{eq:sdot} \frac{d \tilde{\sigma}_a}{d \epsilon} &=& 1 - \frac{1}{q_0} \frac{1}{L_a} \int_0^{L_a} q_a (y) dy ; \\
 \label{eq:xadot} \frac{d \tilde{\chi}_a}{d \epsilon} &=& \kappa_a \, \frac{\tilde{\sigma}_a q}{q_0} \left( 1 - \frac{\tilde{\chi}_a}{\tilde{\chi}_0} \right)  + D_a \, a^2 \, \frac{q_a}{q_0} \frac{\partial^2 \tilde{\chi}_a}{\partial y^2} ; \\
 \label{eq:xcdot} \frac{d \tilde{\chi}_c}{d \epsilon} &=& \kappa_c \sqrt{\tilde{\rho}} \nu (T, \tilde{\rho}, q_0) \left( 1 - \frac{\tilde{\chi}_c}{\tilde{\chi}_0} \right) + D_c \, a^2 \, \frac{\partial^2 \tilde{\chi}_c}{\partial y^2} ; ~~~~~ \\
 \label{eq:rhodot} \frac{d \tilde{\rho}}{d \epsilon} &=& \kappa_1 \frac{\sqrt{\tilde{\rho}}}{\nu (T, \tilde{\rho}, q_0)} \left( 1 - \frac{\tilde{\rho}}{e^{- \beta / \tilde{\chi}_c}} \right) ,
\end{eqnarray}
where $\tilde{\sigma}_a = \sigma_a / E_a$ is the tensile stress in the amorphous layer of width $L_a$ normalized by the Young modulus, and $\beta = e_D / e_Z$. The effective temperatures have been non-dimensionalized by the STZ formation energy $e_Z$: $\tilde{\chi}_a \equiv \chi_A / e_Z$ and $\tilde{\chi}_c \equiv \chi_C / e_Z$. Also, $\kappa_a \equiv E_a / (c_a^{\rm{eff}} e_Z)$ and $\kappa_c \equiv \bar{\mu}_T / (c_c^{\rm{eff}} e_Z)$. We have also dropped the term proportional to $\gamma_D$ in the equation for the effective temperature $\tilde{\chi}_c$ in the crystalline layer, since the results are apparently not sensitive to that term~\cite{langer_2015b}. The plastic strain rate in the amorphous layer is
\begin{equation}\label{eq:a_vpl2}
 q_a = \tau \dot{\epsilon}^{\rm{pl}}_a = \frac{4}{\sqrt{3}} \epsilon_0 \, e^{-1 / \tilde{\chi}_a} \, {\cal C}(\tilde{\sigma}_a) \left( {\cal T}(\tilde{\sigma}_a) - m \right),
\end{equation}
where
\begin{eqnarray}
 {\cal C}(\tilde{\sigma}_a) &=& \exp \left( - \frac{T_E}{T} \right) \cosh \left( \frac{\epsilon_0 \tilde{\sigma}_a}{\sqrt{3} \tilde{e}_Z \tilde{\chi}_a } \right) ; \\
 {\cal T}(\tilde{\sigma}_a) &=& \tanh \left( \frac{\epsilon_0 \tilde{\sigma}_a}{\sqrt{3} \tilde{e}_Z \tilde{\chi}_a} \right) ; \\
 m &=& \cases{
  {\cal T}(\tilde{\sigma}_a) & if $\tilde{\sigma}_a {\cal T}(\tilde{\sigma}_a) \leq \tilde{\sigma}_0$, \\
 \tilde{\sigma}_0 / \tilde{\sigma}_a & if $\tilde{\sigma}_a {\cal T}(\tilde{\sigma}_a) > \tilde{\sigma}_0$ .
 }
\end{eqnarray}
$\tilde{e}_Z$ is the STZ formation energy scaled by $E_a a^3$: $\tilde{e}_Z \equiv e_Z / (E_a a^3)$. Also, the tensile stress in the crystalline layer is directly given by
\begin{equation}
 \sigma_c = \bar{\mu}_T \sqrt{\tilde{\rho}} \nu(T, \tilde{\rho}, q_0 ).
\end{equation}

\section{Model predictions and comparison with experiments}\label{sec:6}

\begin{figure}
\begin{center}
\includegraphics[scale=0.8]{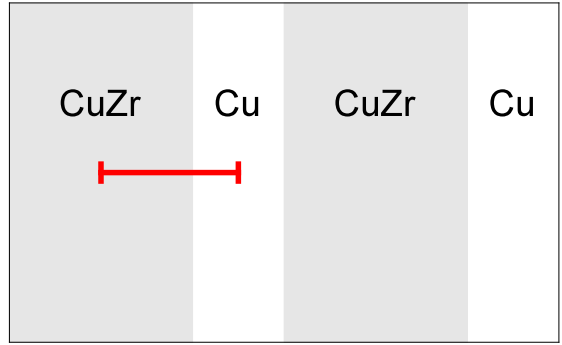}
\caption{\label{fig:schematic}Schematic illustration of the domain numerical solution of the evolution equations that describe the amorphous-crystalline nanolaminate subject to tensile deformation. Assuming symmetry about the center plane of each layer, it suffices to solve the equations on the one-dimensional domain that stretches from the middle of one amorphous CuZr layer to the middle of the adjacent crystalline Cu layer, depicted by the red line.}
\end{center}
\end{figure}

The equations of motion, (\ref{eq:sdot}) through (\ref{eq:rhodot}), are integrated using an adaptive time-stepping scheme based on the Crank-Nicolson method, with uniform spatial discretization (distance between two adjacent grid points is 0.5 nm). Because of symmetry, we confine ourselves to a transverse, one-dimensional domain stretching from the middle of an amorphous CuZr layer to the middle of the adjacent crystalline Cu layer, perpendicularly crossing the ACI. This is illustrated in Figure \ref{fig:schematic}. The initial conditions are $\tilde{\sigma}_a = 10^{-5}$ (small but nonzero to facilitate numerical solution), $\tilde{\chi}_a = \tilde{\chi}_c = 0.032$ across the sample, and $\tilde{\rho} = 10^{-7}$. The thickness of the crystalline Cu layer is fixed at $h_c = 16$ nm, while the amorphous CuZr layer thickness is varied in order to compare to the Kim, Jang and Greer experiment~\cite{kim_2011}. The parameter values are listed in Table \ref{tab:parameters}. Many of these parameter values are documented in the literature (e.g.,~\cite{langer_2015b}), with a few exceptions. For example, with a Young's modulus of $E_a = 72$ GPa for amorphous CuZr inferred from \cite{kim_2011}, and an STZ formation energy $e_Z$ of the order of 1 eV, the dimensionless STZ formation energy roughly equals $\tilde{e}_Z \sim {\cal O}(1)$. Then, $\chi_0 / e_D = 0.25$ according to \cite{langer_2015b}; but because we have chosen $\tilde{\chi}_0 = 0.04$ here, which is roughly consistent with estimates in, for example, \cite{langer_2008}, we choose $\beta = e_D / e_Z = (\chi_0 / e_Z) / (\chi_0 / e_D) = 0.16$. Next, the grain-size-dependent conversion factor $\kappa_1$ that specifies the fraction of energy converted into dislocations was of order ${\cal O}(1)$ in \cite{langer_2015b} for grain sizes of order 10 $\mu$m, and is an increasing function of decreasing grain size. For nanolaminates $\kappa_1$ should be considerably larger, and we have chosen $\kappa_1 = 30$. Finally, we choose for the effective temperature diffusion coefficients $D_a = 10^7$ and $D_c = 1.5 \times 10^4$; our choice stipulates that the diffusion of disorder in the crystalline layer is much slower than in the amorphous layer.

\begin{table}[h]
\caption{\label{tab:parameters}List of variables and parameter values.}
\begin{indented}
\item[]\begin{tabular}{@{}llll}
%\begin{tabular}{p{2cm} p{7.5cm} p{3.5cm}}
\br
Variable & Description & Value \\
\mr
$\tilde{\chi}_0$ & Steady-state effective temperature & 0.04 \\
$E_a$ & Young's modulus of CuZr & 72 GPa \\
$\bar{\mu}_T$ & Effective shear modulus & 10 GPa \\
$T$ & Thermal temperature & 298 K \\
$T_P$ & Depinning temperature & $4.08 \times 10^4$ K \cite{langer_2015b} \\
$T_E$ & STZ activation temperature & 600 K \cite{langer_2008} \\
$\kappa_1$ & Conversion factor & 30 \\
$\kappa_c$ & Conversion factor & 11 \cite{langer_2015b} \\
$\kappa_a$ & Conversion factor & 80 \\ % Should be 80?
$\tau$ & Dimensionless loading rate & $10^{-12}$ s \cite{langer_2008,langer_2015b} \\
$\epsilon_0$ & STZ core volume in units of $a^3$ & 1.5 \\
$\tilde{e}_Z$ & Rescaled STZ formation energy & 1.0 \\
$\beta$ & Dislocation-STZ energy ratio & 0.16 \\
$\tilde{\sigma}_0$ & STZ yield stress parameter & 0.02 \cite{langer_2008} \\
$D_a$ & Diffusion constant in amorphous layer & $10^7$ \\
$D_c$ & Diffusion constant in crystalline layer & $1.5 \times 10^4$ \\
$a$ & Atomic size & 0.167 nm \\
\br
\end{tabular}
\end{indented}
\end{table}

\begin{figure}
\begin{center}
\includegraphics[scale=0.8]{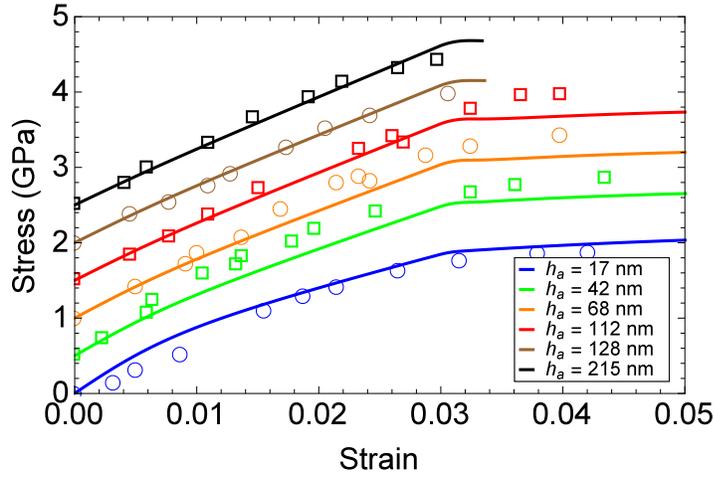}
\caption{\label{fig:plotse}Variation of tensile stress $\sigma$ with strain $\epsilon$, for various values of the amorphous CuZr layer thickness $h_a$. The crystalline layer thickness is $h_c = 16$ nm, and the strain rate is $\dot{\epsilon} = 10^{-3}$ s$^{-1}$. The open circles and squares are the stress levels captured from \cite{kim_2011}. The curves and data points have been offset vertically by 0.5 MPa for each pair of adjacent values of amorphous layer thickness $h_a$ for clarity.}
\end{center}
\end{figure}

\begin{figure}
\begin{center}
\includegraphics[scale=0.8]{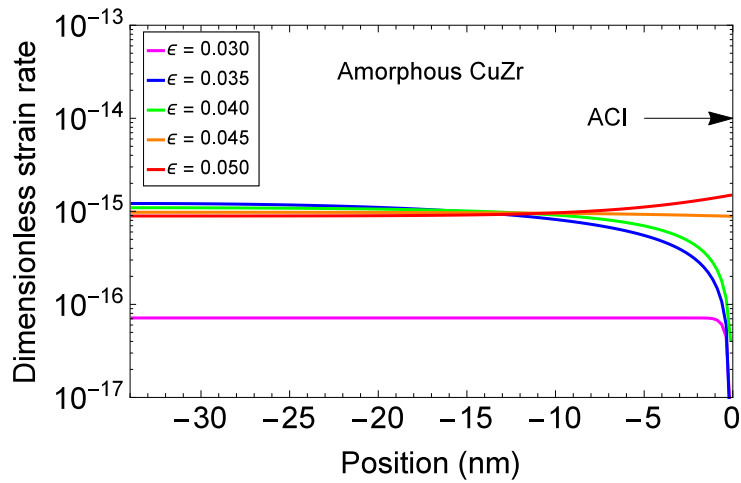}
\caption{\label{fig:qaplot}Variation of dimensionless plastic strain rate $q$ across the half-width of the amorphous CuZr layer, at various snapshots of total accumulated strain $\epsilon$. The amorphous layer thickness is $h_a = 68$ nm, so that $y = -34$ nm is the center axis of the amorphous layer. The applied loading rate is $\dot{\epsilon} = 10^{-3}$ s$^{-1}$; with $\tau = 10^{-12}$ s, the dimensionless loading rate is $q_0 = 10^{-15}$.}
\end{center}
\end{figure}

\begin{figure}
\begin{center}
\includegraphics[scale=0.8]{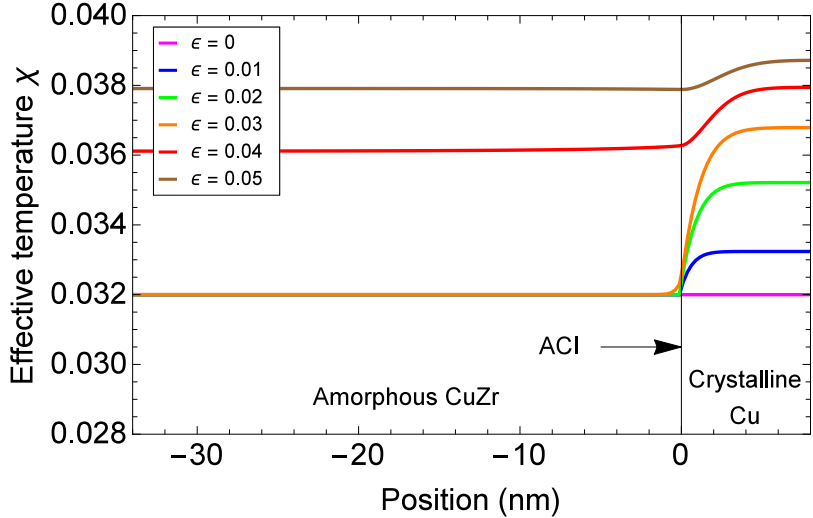}
\caption{\label{fig:xplot}Nondimensionalized effective temperature $\tilde{\chi}$ across the amorphous-crystalline nanolaminate, at various snapshots of total accumulated strain $\epsilon$. Here $\kappa_c = 10$; other parameters are listed in Table \ref{tab:parameters}. The position $y = 0$ is the amorphous-crystalline interface. Here the thickness of the amorphous CuZr layer is $h_a = 68$ nm, while that of the crystalline Cu layer is $h_c = 16$ nm. Thus $y = -34$ nm is the center axis of the amorphous layer, while $y = 8$ nm is the center axis of the crystalline layer. The diffusion of effective temperature and hence configurational disorder is largely determined by the effective temperature gradient across the interface $y = 0$.}
\end{center}
\end{figure}

Figure \ref{fig:plotse} shows the variation of the tensile stress $\sigma$ with the accumulated strain $\epsilon$, for various values of the amorphous layer thickness $h_a$. Our choice of the parameters $\tilde{\sigma}_0$, $E_a$, and $\bar{\mu}_T$ ensures that the heterogeneous material yields at strain $\epsilon \approx 0.03$ and stress $\sigma \approx 2$ GPa, roughly consistent with experiments in \cite{kim_2011}. Notice that the stress-strain curves for $h_a = 128$ and 215 nm break off at strains $\epsilon \sim 0.033$, substantially earlier than the nanolaminates with thinner amorphous layer thickness, in agreement with~\cite{kim_2011}. We compute these curves based on the postulate that failure occurs when the plastic strain rate $\dot{\epsilon}^{\rm{pl}}$ at the edge of the amorphous layer falls off to zero. Indeed, this behavior, dependent on the initial conditions as well as the choice of parameters -- especially $D_a$ -- is seen in the numerical solutions to the equations of motion. Figure \ref{fig:qaplot} shows the strain rate profile at a thickness $h_a = 68$ nm, below the critical thickness for early material failure. The strain rate near the amorphous-crystalline interface is close to zero, at least immediately after the onset of plastic deformation in the amorphous layer, while for $h_a = 68$ nm the strain rate profile quickly becomes more or less uniform. This rapid approach to uniformity may not be the case for nanolaminates with a thicker amorphous layer. This point will be discussed in more detail afterwards, in conjunction with the effective temperature profile shown in figure~\ref{fig:xplot}.

\begin{figure}
\begin{center}
\includegraphics[scale=0.8]{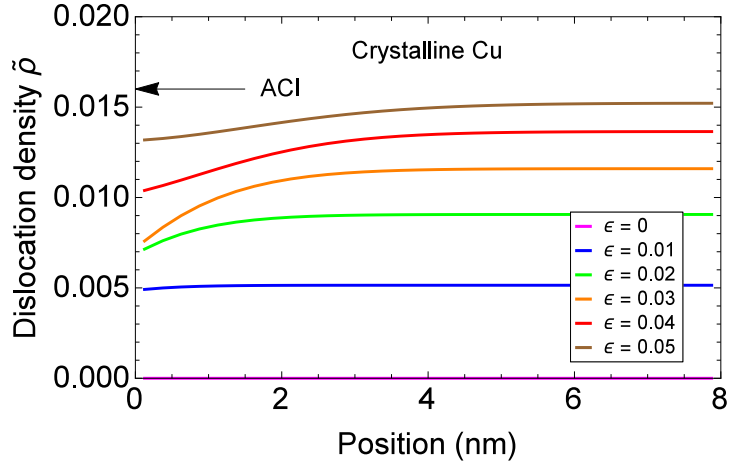}
\caption{\label{fig:rhplot}Nondimensionalized dislocation density $\tilde{\rho}$ across the half-width of the crystalline Cu layer, at various snapshots of total accumulated strain $\epsilon$. Here $\kappa_c = 10$; other parameters are listed in Table \ref{tab:parameters}. The position $y = 0$ is the interface with the amorphous CuZr layer, as indicated by the arrow, while $y = 8$ nm is the center axis of the crystalline layer. As tensile deformation continues the dislocation density increases in the Cu layer, but decreases towards the amorphous-crystalline interface.}
\end{center}
\end{figure}

Figure \ref{fig:rhplot} shows the nondimensionalized dislocation density $\tilde{\rho}$ across the half-width of the crystalline Cu layer at various snapshots of the total accumulated strain $\epsilon$ or, equivalently, time. The dislocation density increases with increasing strain, as it should, and decreases towards the amorphous-crystalline interface at position $y = 0$. Thus our choice of parameters suggests the absorption of dislocations by the interface, in concordance with simulations such as \cite{wang_2007}. It is worth noting that the interface is a more effective sink of dislocations prior to the yielding of the amorphous layer than after.

Finally, Figure \ref{fig:xplot} shows snapshots of the effective temperature distribution across the half-width from the center of the amorphous CuZr layer to the center of the crystalline layer. The effective temperature in the amorphous layer $\tilde{\chi}_a$ remains constant prior to yield ($\epsilon = 0.03$), while it increases in the crystalline layer from the outset of deformation as dislocations are perpetually created. Once the amorphous layer starts to yield, however, diffusion of configurational disorder and hence the effective temperature becomes possible through the interface at $y = 0$ via equation (\ref{eq:xadot}). The direction of diffusion of the effective temperature is largely determined by its gradient across the interface. For the present choice of parameters -- specifically with $\kappa_c = 10$ -- it seems that the effective temperature in the amorphous layer increases more slowly than in the crystalline layer, at least during the early stages of plastic deformation in amorphous CuZr. As such, disorder diffuses from the crystalline layer to the amorphous layer, signalling the absorption of dislocations from the Cu layer into the interface, and the subsequent nucleation of STZs at the interface into the CuZr layer. This is consistent with the experimental observations reported in \cite{wang_2007}. For other parameter choices not shown here -- specifically a larger ratio of  $\kappa_a/\kappa_c$, it is possible for the effective temperature in the amorphous layer to increase faster than in the crystalline layer, setting up an effective temperature gradient across the ACI opposite to the one in the present case. In such a case, the reverse may occur, i.e., the interface would act as a sink of STZs and a source of dislocations, as in \cite{zhang_2013}. Delineation of the exact mechanism is likely material-dependent and requires further microscopic imaging during laboratory studies on a case-by-case basis.

If we compare figures \ref{fig:qaplot} and \ref{fig:xplot}, however, it becomes evident that an increased STZ density through a higher effective temperature does not automatically imply an elevated plastic strain rate. This is a purely entropic effect. To understand why this happens, recall from (\ref{eq:a_vpl2}) that the effective temperature $\tilde{\chi}_a$ controls the plastic strain rate not just through the STZ density $\Lambda = 2 e^{-1 / \tilde{\chi}_a}$, but also through the rate factors ${\cal C}(\tilde{\sigma}_a) \propto \cosh [\epsilon_0 \tilde{\sigma}_a / (\sqrt{3} \tilde{e}_Z \tilde{\chi}_a)]$, and ${\cal T}(\tilde{\sigma}_a) = \tanh [\epsilon_0 \tilde{\sigma}_a / (\sqrt{3} \tilde{e}_Z \tilde{\chi}_a)]$. The argument of these hyperbolic trigonometric functions is a decreasing function of increasing effective temperature $\tilde{\chi}_a$ in the amorphous layer. While the STZ density is an increasing function of $\tilde{\chi}_a$, there is a range of $\tilde{\chi}_a$ over which $q$ decreases as a function of increasing $\tilde{\chi}_a$. Physically, while the effective temperature near the edge of the amorphous layer increases as a result of effective heat transfer -- or diffusion of disorder -- from the crystalline layer, the STZs produced at the interface do not contribute to plastic strain until they move deeper into the amorphous layer. Importantly, if $\tilde{\chi}_a$ is large enough, the plastic strain rate goes to zero since $\tilde{\sigma}_a {\cal T}(\tilde{\sigma}_a) < \tilde{\sigma}_0$ such that ${\cal T}(\tilde{\sigma}_a) - m = 0$. The amorphous material near the interface becomes so disordered that the applied stress can no longer sustain the strain and the material fails.

\section{Summary and concluding remarks}

In this paper, we presented an effective-temperature framework that statistically describes the motion of and interaction between plasticity carriers (dislocations in the crystalline layers, and STZs in the amorphous layers) across an ACI in a natural manner. The effective temperature controls the dynamics of defects in a deforming solid, and describes the slow, configurational degrees of freedom that correspond to the infrequent atomic rearrangements associated with irreversible plastic deformation. The absorption of plasticity carriers on one side of the ACI and the subsequent production of plasticity carriers that move deep into the other side is interpreted as the flow of configurational disorder across the interface. Given our choice of parameters, we find the ACI to be a sink of dislocations in the crystalline Cu layer and a source of STZs that move into amorphous CuZr, as observed in experiments and simulations such as ~\cite{wang_2007}. In addition, we have demonstrated the direct link between effective-temperature diffusion and the size-dependent ultimate tensile strength of the heterogeneous multilayered nanolaminate structure. Crucially, the effective-temperature theories of dislocations in crystalline solids, and STZs in amorphous solids, are fully consistent with the principles of energy conservation, nondecreasing entropy, and symmetry. With only a small handful of equations and parameters we have been able to obtain reasonably good fits to experiments of nanolaminates under tensile deformation.

The theory presented here describes the dynamics of STZs in metallic glasses and dislocations in nanocrystalline materials with a single effective temperature. It opens up new avenues for describing co-deformation in heterogeneous structures under different loading conditions such as uniaxial compression and shear~\cite{arman_2011,brandl_2013}. In the present case of amorphous-crystalline nanolaminates, further microsopy studies and molecular-dynamics simulations may serve to illuminate the direction in which plasticity carriers move and configurational disorder flows across the interface between the two different structures. Nonetheless, the present effective-temperature approach serves to provide a simple and unified description of interacting plasticity carriers. It is unclear to us how one can combine traditional dislocation theories with other theories of flow in metallic glasses, without adding extra empirical, if not unphysical, equations, that make additional assumptions about the interaction between plasticity carriers on the two sides of the ACI.

\section*{Acknowledgments}

CL was partially funded by the Center for Nonlinear Studies at the Los Alamos National Laboratory over the course of this work. JRM acknowledges the support of the Los Alamos National Laboratory Directed Research and Development (LDRD) Early Career Award 20150696ECR.

\clearpage

\appendix

\bibliographystyle{iopart-num}
\bibliography{msmse_aci_01}

\end{document}